# Mapping the Energetic Structure of Climate Transitions for Policy Relevant Regime Detection

**Ngueuleweu Tiwang Gildas(Ph.D)**

Email: *gildas.ngueuleweu@ univ-dschang.org*

**Abstract:**

Understanding how climate and innovation policies perform during socio-technical transitions remains a central challenge in innovation studies. Empirical analyses of the relationship between economic growth and carbon emissions continue to yield conflicting results, partly because they rely on pooled models that implicitly assume stable and homogeneous dynamics. Transition theory, by contrast, emphasizes that decarbonization unfolds through heterogeneous regimes characterized by varying degrees of stability, inertia, and reconfiguration. Yet, empirical tools capable of identifying these regimes prior to policy evaluation or forecasting remain limited. This paper introduces a regime-diagnostic framework designed to condition empirical analysis on the structural state of the climat-economy system. Rather than estimating causal effects or generating forecasts directly, the framework reconstructs latent transition regimes from the time varying responsiveness of emissions to economic activity. These diagnostics are used as a pre-modeling step, allowing econometric and machine learning tools to be applied conditionally on empirically identified regimes. Using a global panel of approximately 150 countries over the period 1991-2022, we apply the framework to the $CO_2$-GDP relationship. The results reveal three recurrent regimes, flow-dominant, transitional, and store-dominant, each associated with distinct patterns of predictability and model performance. We show that forecasting accuracy, causal inference, and early warning signals vary systematically across regimes. In particular, transitional regimes exhibit heightened instability and weaker policy-relevant signal quality, while structurally stable regimes support more reliable empirical inference. The findings contribute to the innovation and transition policy literature in three ways. First, they provide empirical evidence that climate-growth dynamics are regime dependent rather than governed by a single development trajectory. Second, they demonstrate that policy evaluation outcomes depend critically on the transition state in which policies are deployed. Third, they advance a regime-aware empirical approach that enhances policy learning by reducing inference errors arising from pooled estimation. Overall, the study highlights the importance of diagnosing structural conditions prior to policy analysis and offers a foundation for more adaptive, context-sensitive innovation and transition governance.

# 1. Introduction

Achieving climate mitigation while sustaining economic growth remains a central challenge for innovation and transition policy. Despite decades of empirical research, the relationship between economic development and carbon emissions continues to yield conflicting conclusions. While early work suggested the possibility of eventual decoupling between income growth and environmental pressure (Grossman & Krueger, 1995), subsequent studies document persistent coupling, recoupling, and highly heterogeneous trajectories across countries and periods (Stern, 2004; Wiedmann et al., 2015; Stern, 2017). This lack of consensus has constrained policy learning: climate, industrial, and innovation policies are often designed, evaluated, and compared under the implicit assumption that economic-environmental relationships are stable and globally valid. When this assumption fails, policy evaluations risk attributing success or failure to policy design rather than to underlying structural conditions.Recent advances in innovation and transition studies challenge this assumption by emphasizing that decarbonization unfolds through heterogeneous, path-dependent regimes embedded in evolving socio-technical systems (Unruh, 2000; Geels, 2011; Köhler et al., 2019). From this perspective, innovation policy effectiveness depends not only on instrument choice but also on policy paradigms (Schot & Steinmueller, 2018), directionality and mission orientation (Mazzucato, 2018), and the interaction between growth dynamics and policy design (Aghion et al., 2016). Mission-oriented policies, carbon pricing schemes, and green industrial strategies operate within regimes characterized by technological lock-in, institutional inertia, and feedback effects. As a result, identical policy instruments may generate sharply different outcomes depending on the phase of transition in which they are deployed.

This regime dependence poses a fundamental challenge for policy evaluation and learning, a core concern of Research Policy. Much of the existing empirical literature evaluates climate and innovation policies using pooled models that implicitly assume a single underlying regime. Such approaches obscure structural heterogeneity and weaken policy inference, making it difficult to distinguish genuine policy effects from regime-driven dynamics. In turn, this undermines cumulative learning, cross-country comparison, and evidence-based policy design. While transition theory highlights the importance of regimes, empirical tools capable of diagnosing them in a policy-relevant and forward-looking manner remain limited. Existing approaches face important constraints. Structural break tests, threshold models, and Markov switching frameworks typically identify regime shifts retrospectively and rely on strong ex-ante assumptions about regime structure (Hamilton, 1989; Hansen, 2000; Bai & Perron, 2003). Machine learning methods can uncover latent patterns but often lack interpretability and explicit links to policy mechanisms, limiting their usefulness for policy learning and governance (Reichstein et al., 2019). Early warning indicators derived from critical transition theory provide insights into instability, yet they are highly sensitive to noise and are rarely integrated into systematic innovation policy evaluation (Scheffer et al., 2009; Dakos et al., 2012). As a consequence, policy assessments are frequently conducted without a clear understanding of whether the underlying system is in a structurally stable or unstable state—precisely when evaluation results are most likely to be misleading.

To address this gap, this paper introduces a regime-diagnostic policy tool designed to enhance the reliability of innovation and transition policy analysis. Rather than estimating causal effects or producing forecasts directly, the proposed framework reconstructs latent transition regimes from the endogenous evolution of the economic-environmental relationship itself. The central premise is that the responsiveness of emissions to economic activity, and critically, the higher order dynamics of this responsiveness contains actionable information about the structural state of the transition. When this information is ignored, policy evaluations may conflate fundamentally different regimes, generating unstable estimates and spurious conclusions about policy effectiveness. We operationalize this diagnostic through the Ngueuleweu Elasticity Energetics Diagnostics (NEED) framework. NEED tracks the time varying elasticity between economic output and emissions and its higher order dynamics to construct interpretable indicators of structural tension, adjustment intensity, and instability within the climate-economy system. Importantly, NEED is not proposed as a substitute for econometric or machine learning models. Instead, it functions as a pre-modeling policy diagnostic that conditions when and how such models should be applied. In this sense, NEED directly addresses a central concern in policy evaluation and learning: identifying when policy estimates and forecasts are structurally reliable, and when they are not.

Empirically, we apply NEED to a global panel of approximately 150 countries over the period 1991-2022, focusing on the $CO_2$-GDP relationship. The analysis reveals three recurrent regimes(flow-dominant, transitional, and store-dominant), each associated with distinct patterns of adjustment, stability, and predictability. We show that forecasting accuracy, causal inference, and early warning performance vary sharply across these regimes. Policy relevant signals are most reliable in structurally stable regimes, while transitional phases exhibit heightened volatility and degraded model performance. These results provide a structural explanation for why similar climate and innovation policies often yield divergent outcomes across countries and periods, consistent with regime-based transition theories (Unruh, 2000; Geels, 2011; Köhler et al., 2019).

This study contributes to the Research Policy literature in three ways. First, it introduces a regime-aware policy diagnostic that enhances innovation and transition policy evaluation by identifying when empirical estimates and forecasts are structurally trustworthy. Second, it provides global evidence that climate-growth dynamics are governed by regime-dependent processes rather than a single universal trajectory, extending debates on heterogeneous decarbonization pathways and the limits of average effect analysis (Stern, 2017; Wiedmann et al., 2015). Third, it advances empirical practice in innovation policy research by embedding econometric and machine learning analyses within empirically identified regimes, thereby reducing false policy inferences arising from pooled estimation. More broadly, by shifting attention from average relationships to regime, conditioned policy reliability, this paper contributes to ongoing debates on policy timing, sequencing, and adaptive transition governance. The findings suggest that the effectiveness of climate and innovation policies depends less on their intrinsic design than on the structural state of the system in which they are deployed. Diagnosing these states is therefore not an academic exercise, but a prerequisite for credible policy evaluation and effective mission-oriented innovation policy.Importantly, NEED does not aim to tell policymakers which policies to adopt. Instead, it identifies when policy evaluation and empirical inference are structurally informative, and when they are likely to be misleading due to regime instability.

# 2. Methodology

## 2.1. Empirical motivation and identification challenge

Empirical analyses of climate-economy relationships typically rely on pooled models that assume a stable underlying structure linking economic activity and emissions. However, transition theory and innovation studies emphasize that socio-technical systems evolve through heterogeneous regimes characterized by different adjustment dynamics, degrees of inertia, and policy responsiveness (Unruh, 2000; Geels, 2011; Markard, 2018). When such regimes are ignored, pooled estimation may conflate structurally distinct phases, leading to unstable coefficients, misleading forecasts, and weak policy inference. The methodological challenge addressed in this paper is therefore not to estimate a new causal model of emissions, but to identify when standard empirical tools, econometric or machine learning based, are structurally reliable. To this end, we adopt a regime diagnostic approach that reconstructs latent transition states prior to forecasting, causal inference, or early warning analysis.

## 2.2. Measuring time-varying responsiveness

We begin by measuring the time varying responsiveness of carbon emissions to economic activity using elasticity. For each country $i$ and year $t$, elasticity is defined as the log-log derivative of emissions with respect to GDP:

$$\varepsilon_t = \frac{\mathrm{d} \ln Y_t}{\mathrm{d} \ln X_t} \tag{1}$$

Elasticities are estimated using rolling windows to capture gradual structural change while preserving sufficient local variation. This approach allows the emissions-growth relationship to evolve over time rather than imposing a constant parameter across heterogeneous development phases. Because rolling estimates and their temporal changes are sensitive to noise, especially in long macroeconomic panels, we apply a state-space smoothing procedure to obtain stable estimates of elasticity and its temporal dynamics. This step improves signal extraction without altering the underlying data generating process.

## 2.3. Constructing regime diagnostics

To capture how the emissions-growth relationship evolves beyond its level, we track changes in elasticity over time. These dynamics reflect whether the system is stabilizing, adjusting smoothly, or entering phases of heightened instability. Rather than interpreting these dynamics as causal forces, we treat them as descriptive indicators of structural state.

From the smoothed elasticity series, we derive a set of indicators summarizing: (i) the intensity of structural tension between emissions and growth, (ii) the pace of adjustment of this relationship, and (iii) the degree of instability associated with abrupt changes.

These indicators are combined into a compact diagnostic representation of the system's structural condition at each point in time. Importantly, this diagnostic is used exclusively to identify regimes and does not enter subsequent estimation equations as explanatory variables.

## 2.4. Endogenous regime identification

Latent regimes are identified endogenously using clustering techniques applied to the diagnostic indicators. This approach avoids imposing ex ante thresholds or assuming a fixed number of transition phases. Instead, regimes emerge from the joint distribution of structural tension and adjustment dynamics observed in the data.

The resulting classification consistently identifies three broad regimes across countries and time periods. The first corresponds to highly reactive, low inertia dynamics, where emissions respond strongly to economic activity. The second reflects transitional phases characterized by heightened volatility and structural reconfiguration. The third corresponds to high inertia regimes in which emissions-growth relationships stabilize but adjust slowly.These regimes are interpreted as empirical representations of distinct transition states rather than as equilibrium outcomes.

## 2.5. Regime-conditioned empirical strategy

Once regimes are identified, all subsequent analyses are conducted conditionally on regime membership. Specifically, we re-estimate forecasting models, causal relationships, and early warning indicators separately within each regime. This strategy allows us to evaluate how model performance and inference vary across structural states. We compare regime conditioned results with pooled estimates to assess the extent to which ignoring regime heterogeneity obscures policy relevant dynamics. Model performance is evaluated using out-of-sample accuracy, classification metrics, and lead time diagnostics, with particular attention to differences between stable and transitional regimes. By construction, this approach does not privilege any specific econometric or machine learning model. Instead, it provides a diagnostic framework within which existing tools can be applied more reliably.

## 2.6. Data and implementation

The empirical analysis uses a global panel of approximately 150 countries covering the period 1991-2022. Carbon dioxide emissions and GDP data are drawn from the World Development Indicators database. All variables are transformed into logarithms prior to elasticity estimation. Rolling window estimation, smoothing procedures, regime identification, and subsequent analyses are implemented uniformly across countries to ensure comparability. Additional robustness checks, alternative clustering specifications, and technical details are reported in Appendix.

## 2.7. Interpretation and scope

The proposed methodology is not intended to replace structural models of innovation, growth, or emissions. Rather, it provides a pre-modeling diagnostic that conditions empirical analysis on the system's transition state. By identifying when relationships are stable, unstable, or reconfiguring, the framework clarifies when policy evaluation and forecasting are likely to be informative and when they are not.

In this sense, the methodology contributes to innovation and transition studies by operationalizing regime awareness in empirical policy analysis, without imposing strong theoretical assumptions or restricting the choice of downstream models.

# 3. Results and Discussions

This section presents NEED framework implementation. Results are compared at three levels: first, analyses of NEED per subregion(annex), then we define 3 regimes exogenously from a simple 1/3 sample rule based on ascending values of indicators, and finally, we allow NEED to detect regimes automatically. Analyses are compared by scenario using several models. Emphasis is placed on stability, performance, and risk per scenario, as well as their consistency with the theoretical expectations of the NEED model. The objective is to investigate the differences per scenario of NEED on prediction, causality, early warning, and any other analyses.

## 3.1. Performance of NEED within defined structural states

Table 1 identifies store, transitional and flow dominant energies which share the sample in 3 parts for new analyses. It demonstrates that forecasting accuracy in the climate-economy system varies markedly across energetic regimes. This confirms the structural heterogeneity hypothesized in transition studies (Geels, 2011; Köhler et al., 2019). While global models(the total sample of study) achieve high out-of-sample fit ($R^2 \approx 0.98$), regime specific forecasts, particularly in store dominant phases, perform substantially better ($R^2 = 0.992\text{-}0.993$). This result reveals that transitions contain distinct and more predictable internal dynamics. This supports the idea that socio-technical systems exhibit phase dependent behaviors, and this is consistent with work on nonlinear climate-economy coupling (Lenton et al., 2019). By contrast, transitional regimes show the weakest predictive performance across all models. This can be aligned with literature on transition phases as periods of heightened uncertainty and structural reconfiguration (Markard, 2018). These findings empirically validate the NEED framework's claim that forecasting is fundamentally regime sensitive rather than globally uniform.

**Table 1: Causality performance of NEED within regimes defined exogenously**,

| Regime | Model | RMSE | MAE | R2_OOS | n_test |
|---|---|---|---|---|---|
| Global | OLS | 0.153 | 0.064 | 0.985 | 1380 |
| Global | RF | 0.183 | 0.076 | 0.979 | 1380 |
| Global | VAR(1) | 0.179 | 0.079 | 0.98 | 1380 |
| Global | XGB | 0.181 | 0.08 | 0.98 | 1380 |
| Store-dominant | OLS | 0.103 | 0.06 | 0.993 | 227 |
| Store-dominant | RF | 0.111 | 0.067 | 0.992 | 227 |
| Store-dominant | VAR(1) | 0.124 | 0.069 | 0.99 | 227 |
| Store-dominant | XGB | 0.147 | 0.086 | 0.986 | 227 |
| Transitional | OLS | 0.198 | 0.078 | 0.975 | 444 |
| Transitional | RF | 0.221 | 0.088 | 0.969 | 444 |
| Transitional | VAR(1) | 0.206 | 0.082 | 0.973 | 444 |
| Transitional | XGB | 0.241 | 0.111 | 0.963 | 444 |
| Flow-dominant | OLS | 0.155 | 0.07 | 0.986 | 479 |
| Flow-dominant | RF | 0.186 | 0.078 | 0.98 | 479 |
| Flow-dominant | VAR(1) | 0.181 | 0.076 | 0.981 | 479 |
| Flow-dominant | XGB | 0.21 | 0.107 | 0.975 | 479 |

These results of table 1 extend the $CO_2$-GDP debate on the influence of decoupling potential and trajectory predictability on a country's energetic regime, rather than on global averages implied by EKC-type models (Stern, 2017; Apergis & Payne, 2010). The superior performance of OLS and VAR models within store dominant and flow dominant regimes suggests that structural stability enables reliable emission-productivity forecasting, while the degraded accuracy in transitional regimes highlights when policy interventions face the greatest uncertainty and require adaptive governance (Victor et al., 2019). This regime sensitive predictability complements current discussions on heterogeneous decarbonization pathways (Mattioli et al., 2020).They demonstrate that transitions themselves alter model performance. Thus, NEED is an innovative diagnostic for policymakers: forecasts should be conditioned on the current energetic state of the system, enabling more precise timing of climate, innovation, and industrial strategies. let us check how it performs for prediction in figure 1.

figure 1 confirms that regime specific transition detection is feasible even under externally imposed segmentation. Ensemble models(Random Forest and XGBoost) reliably outperform kurtosis and skewness based thresholds in both flow dominant and store dominant regimes, with PR-lift values as high as 8.8-11.0. This indicates a strong signal to noise amplification (Breiman, 2001; Chen & Guestrin, 2016). This is consistent with research on lock in and stabilizing feedbacks in socio-technical systems (Unruh, 2000). In contrast, transitional regimes exhibit the lowest predictive lift and weakest F1 scores across all models.This aligns with Markard(2018) justification that transition phases are characterized by elevated volatility and structural reconfiguration. These results complement earlier regime learning findings by showing that the underlying energetic structure is detectable even without automatic identification, underscoring the robustness of NEED as a diagnostic framework.

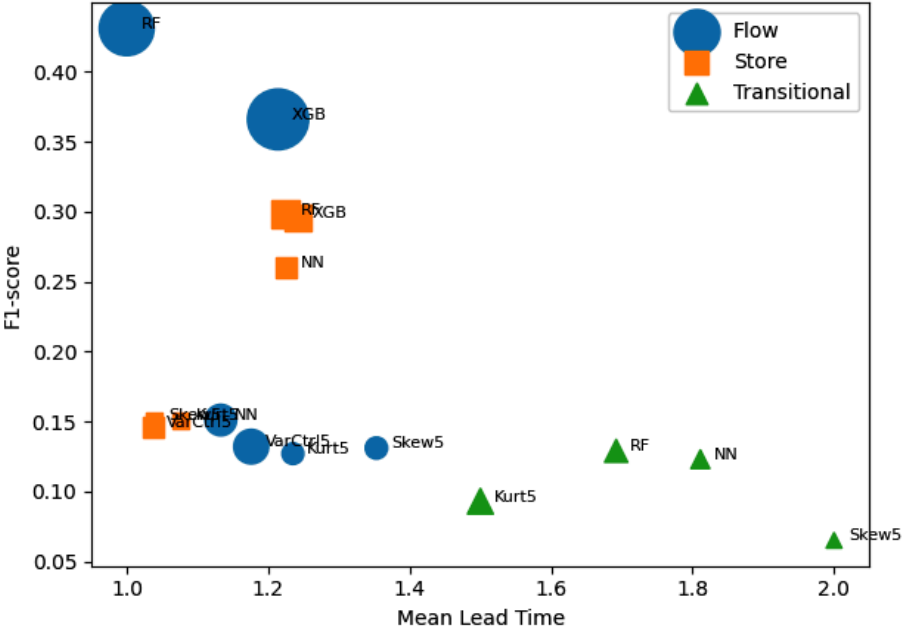

Figure 1: **NEED Regime dignosis: trade-off between early detection and reliability**

These findings of figure 1 demonstrates that the predictability of climate-economic dynamics fundamentally depends on the energetic phase in which a country or region resides. Some regimes deliver high F1 and long predictive lead times.This result implies that mitigation trajectories and decoupling potentials can be anticipated more reliably during stable phases. This insight challenges Stern(2017) or Dinda(2004) global linear assumptions embedded in Envi-ronmental Kuztnet Curve(EKC)-style models. Conversely, the poor performance observed in transitional regimes highlights periods in which policymaking should be especially adaptive and precautionary. This aligns with Victor et al.(2019) call for anticipatory climate governance under uncertainty. By demonstrating that imposed regimes still exhibit coherent machine learnable structures, this analysis strengthens the case for regime aware forecasting tools that can guide the timing of carbon regulation, innovation support, and investment in low-carbon technologies. NEED thus provides a novel computational foundation for identifying when structural conditions are favorable, or unfavorable, for effective climate policy intervention. Why not allow automtic regime selection and see how it affects outcome.

### 3.2. Performance of NEED within automatic detected regimes

Table 2 shows that when energetic regimes are identified automatically rather than imposed, the global climate-economy system overwhelmingly allocates observations to the transitional regime (2,077 cases), followed by the store dominant regime (1,484) and very few flow dominant cases (15). This distribution reveals that countries spend most of their developmental trajectory in structurally unstable zones. Here, tensions between emissions and economic activity are actively reconfiguring. This is consistent with Geels(2011) or Markard(2018) transition studies emphasizing prolonged phases of turbulence rather than short discrete shifts. The scarcity of flow dominant cases across all regions indicates that true decoupling remains rare, reinforcing recent findings that stable low carbon productivity regimes have yet to consolidate in most economies (Rogelj et al., 2021). At the same time, the presence of store dominant regimes across East Asia, Europe & Central Asia, and South Asia suggests pockets of structural stability where coordinated policy, innovation, and investment may be more effective. Automatic detection therefore confirms that NEED is sensitive to real world energetic structures rather than to assumptions imposed ex ante.

Table 2: Subregions Automatic Regime Selection,

| subregion | Flow-dominant | Store-dominant | Transitional |
|---|---|---|---|
| East Asia & Pacific | 1 | 6 | 449 |
| Europe & Central Asia | 1 | 933 | 2 |
| Latin America & Caribbean | 5 | 1 | 618 |
| Middle East & North Africa (MENA) | 3 | 323 | 10 |
| North America | 1 | 63 | 8 |
| South Asia | 1 | 152 | 15 |
| Sub-Saharan Africa (SSA) | 3 | 6 | 975 |
| total | 15 | 1484 | 2077 |

These results of table 2 extend the $CO_2$-GDP debate by showing that the likelihood of observing decoupling or re-coupling dynamics depends on the endogenous regime distribution of each region, rather than on global averages or uniform developmental trajectories (Stern, 2017; Burke et al., 2015). The dominance of transitional regimes underscores why empirical studies often report inconsistent or nonlinear decarbonization patterns: most countries are operating in inherently unstable energetic states where standard econometric relationships break down. Conversely, the clustering of store dominant observations in selected regions suggests the existence of policy windows, which for Kemp et al.(2007) are periods when structural conditions temporarily stabilize and climate or innovation policies can achieve disproportionately large effects. The near absence of flow dominant regimes highlights how far the global economy remains from sustained decoupling. This result offers a strong empirical support for the argument that decarbonization requires deliberate, regime aware interventions rather than spontaneous market adjustments. NEED, therefore, brings a novel diagnostic capability to transition governance by identifying when and where structural conditions are aligned or misaligned with effective climate policy. An analysis under these regimes can be informative(table 3).

Table 3 shows that once regimes are identified automatically, machine learning models, especially Random Forests and XGBoost, are able to recover the structure of both Regime 2 and Regime 3 with substantial precision. The strong AUPRC values (0.59-0.62 for Regime 2; 0.45-0.58 for Regime 3) and high PR-lift levels, particularly for Regime 3 (17-22 above baseline), indicate that these energetic configurations possess robust and detectable signatures rather than random fluctuations. This reinforces the idea that transitions in the climate-economy system exhibit latent structural regularities, consistent with complexity based transition theories (Geels, 2011; Arthur, 2021). The weaker performance of classical variance control approaches underscores the inadequacy of linear or Gaussian assumptions in capturing these dynamics. This complement recent work, showing that socio-technical transitions involve nonlinear tipping elements and emergent feedbacks (Lenton et al., 2019; Farmer et al., 2019). Collectively, these findings confirm that NEED's regime signals can be learned directly from data without externally imposed segmentation.

**Table 3: Prediction Performance within Automatic Regime Selection**,

| Regime | Model | AUPRC | AU-ROC | Base rate | PR Lift | Precision | Recall | F1 | Lead time mean | Detect Rate |
|---|---|---|---|---|---|---|---|---|---|---|
| Regime2 | RF | 0.592 | 0.867 | 0.097 | 6.131 | 0.506 | 0.576 | 0.539 | 1.304 | 0.637 |
| Regime3 | RF | 0.578 | 0.808 | 0.026 | 22.213 | 0.333 | 0.571 | 0.421 | 1 | 0.444 |
| Regime2 | VarCtrl5 | 0.172 | 0.492 | 0.097 | 1.78 | 0.103 | 0.8 | 0.182 | 1.21 | 0.877 |
| Regime3 | VarCtrl5 | 0.056 | 0.546 | 0.026 | 2.134 | 0.031 | 0.629 | 0.059 | 1.636 | 0.611 |
| Regime2 | XGB | 0.617 | 0.877 | 0.097 | 6.383 | 0.566 | 0.645 | 0.603 | 1.102 | 0.693 |
| Regime3 | XGB | 0.454 | 0.778 | 0.026 | 17.443 | 0.303 | 0.571 | 0.396 | 1.111 | 0.5 |

These results(table 3) introduce a new perspective on the $CO_2$-GDP transition debate by showing that different energetic regimes vary not only in frequency but also in predictability and detectability, implying that policy effectiveness is inherently regime dependent. High recall and lead-time performance for both Regime 2 and Regime 3 indicate that transitions into and out of unstable states can be forecasted with meaningful advance warning, an important contribution to anticipatory climate governance (Victor et al., 2019). The exceptionally high PR-lift for Regime 3 suggests that deep transitional phases generate distinctive signals, explaining why empirical studies often report abrupt, nonlinear changes in decarbonization trajectories (Peters et al., 2017). This challenges the global EKC narrative (Stern, 2017) as it demonstrates that emission-income relationships depend on structural energetic states, not uniform developmental paths. NEED therefore provides policymakers with a diagnostic tool capable of identifying when a country is entering a vulnerability window where targeted innovation support, carbon regulation, or industrial transformation may be most effective.

# 4. Implications for Innovation and Transition Policy

## 4.1. Regime dependence and the limits of uniform policy design

A central implication of the findings is that the effectiveness of innovation and climate policies is fundamentally regime dependent. Much of the empirical innovation policy literature implicitly assumes that policy instruments such as carbon pricing, green subsidies, or mission oriented programs, operate on relatively stable structural relationships between economic activity and environmental outcomes. This assumption underpins the widespread use of pooled evaluation frameworks and cross-country policy comparisons. Transition studies, however, have long argued that socio-technical systems evolve through distinct phases of lock-in, destabilization, and reconfiguration, during which policy leverage, feedback mechanisms, and actor responses differ markedly (Unruh, 2000; Geels, 2011; Köhler et al., 2019). The regime diagnostics identified in this paper provide empirical support for this perspective by showing that the climate-economy relationship itself undergoes structurally distinct regimes with different levels of stability and predictability. In structurally stable regimes, policy-relevant signals are more consistent and empirical models exhibit higher reliability, allowing conventional policy instruments to be evaluated with greater confidence. By contrast, transitional regimes, where most countries are empirically observed to reside, are characterized by heightened volatility, unstable elasticities, and degraded model performance. Under such conditions, uniform policy prescriptions applied across heterogeneous regimes are likely to yield uneven or even counterproductive outcomes. This helps explain the mixed and often contradictory evidence on the effectiveness of climate and innovation policies reported in the literature (Stern, 2017; Victor et al., 2019), and underscores the limits of one-size-fits-all policy design.

## 4.2. Policy timing, transition windows, and diagnostic governance

The results also speak directly to debates on policy timing and transition governance. Prior work has emphasized the existence of "windows of opportunity" during which institutional, technological, and behavioral change becomes more feasible (Kemp et al., 2007; Markard, 2018). Yet identifying such windows empirically, and distinguishing them from periods of heightened instability, has remained a persistent challenge. By distinguishing between flow-dominant, transitional, and store-dominant regimes, the diagnostic framework developed in this study offers an operational approach to identifying transition windows using observed data. Store-dominant regimes, characterized by higher inertia but greater structural stability, appear to provide conditions under which policy interventions generate more reliable and interpretable outcomes. Transitional regimes, while often associated with political momentum and heightened expectations, exhibit lower predictability and greater uncertainty, implying higher risks for policy evaluation and learning. Importantly, this does not imply that policy intervention should be postponed during transitional phases. Rather, it suggests that policy instruments must be adapted to regime conditions. During periods of instability, adaptive, experimental, and portfolio-based policy approaches may be more appropriate than rigid, rule-based instruments. This insight aligns closely with calls for reflexive and adaptive governance in transition studies, emphasizing experimentation, feedback, and learning rather than optimization under assumed stability (Geels, 2011; Markard et al., 2012).

## 4.3. Innovation policy, lock-in, and structural inertia

The identification of store-dominant regimes has direct implications for innovation policy in contexts characterized by technological lock-in and structural inertia. A substantial body of research highlights how entrenched production systems, infrastructures, and institutional arrangements can slow the diffusion of low-carbon technologies even when economic incentives are present (Unruh, 2000; Arthur, 2021). The results here suggest that such lock-in corresponds empirically to regimes with high structural inertia but comparatively stable dynamics. In these regimes, incremental policy instruments, such as marginal price signals or isolated subsidies, may be insufficient to induce transformative change. Instead, mission-oriented innovation policies, coordinated public-private investment strategies, and institutional reforms may be required to redirect innovation trajectories and overcome path dependence (Mazzucato, 2018; Köhler et al., 2019). The regime diagnostics developed in this paper help clarify when such intensive interventions are likely to be necessary and when more conventional policy tools may be effective, thereby improving the strategic targeting of innovation policy.

## 4.4. Implications for policy evaluation and learning

A further implication concerns the evaluation of innovation and climate policies, a core concern in the Research Policy literature. Much of the empirical policy evaluation work relies on pooled estimates that average across structurally distinct regimes. The findings of this study demonstrate that such approaches risk attributing policy success or failure to the instrument itself rather than to the transition regime in which it was deployed. Conditioning policy evaluation on empirically identified regimes can substantially improve policy learning by separating genuine policy effects from regime-driven dynamics. In this sense, NEED functions as a policy evaluation reliability diagnostic, identifying when empirical inference is structurally informative and when it is likely to be misleading. This perspective aligns with recent calls for context-sensitive evaluation frameworks that account for system dynamics, feedbacks, and structural change rather than relying on static benchmarks or average effects (Schot & Steinmueller, 2018; Victor et al., 2019).

## 4.5. Forward-looking perspectives for regime-aware transition governance

More broadly, the findings suggest a shift in how innovation and transition policy should be conceptualized and governed. Rather than focusing exclusively on selecting the "right" policy instruments, policymakers and analysts should pay greater attention to diagnosing the structural state of the system prior to intervention. Regime-aware diagnostics can inform not only which policies to deploy, but also when, where, and with what degree of flexibility. From this perspective, the contribution of this paper is not to advocate a specific policy mix, but to provide an empirical foundation for regime-aware transition governance. By embedding policy analysis within empirically identified transition states, future research can move beyond debates over average policy effectiveness toward a more nuanced understanding of policy sequencing, timing, and adaptability in complex socio-technical transitions. This shift is essential for advancing innovation policy frameworks that are capable of learning under uncertainty and governing transitions in real time.

## 4.6. Policy instruments and regime risk

An additional implication of the findings concerns the interaction between policy instruments and regime-specific risk. Innovation and climate policy instruments differ not only in their objectives and mechanisms, but also in their sensitivity to structural instability. Yet most empirical evaluations implicitly assume that policy risk is homogeneous across contexts, treating observed variability in outcomes as noise or implementation failure rather than as a function of underlying transition regimes.The regime diagnostics developed in this study suggest that policy instruments are exposed to systematically different levels of regime risk depending on the structural state of the system. In structurally stable regimes, such as store-dominant phases, policy instruments operate in environments characterized by relatively predictable responses and lower volatility. By contrast, transitional regimes are associated with elevated regime risk. High volatility, shifting elasticities, and unstable feedbacks imply that policy outcomes are more uncertain and that empirical estimates are more fragile. In such contexts, rigid policy instruments that rely on stable behavioral responses may underperform or generate misleading evaluation signals. From this perspective, NEED can be interpreted as a policy risk diagnostic that complements existing policy design frameworks. By identifying the regime in which a policy is deployed, the diagnostic helps distinguish between instrument risk (arising from design or implementation flaws) and regime risk (arising from structural instability). This distinction is critical for policy learning, as it prevents premature abandonment of potentially effective instruments and supports more nuanced adaptation strategies.

## 5. Conclusion

This study has addressed a central challenge in innovation and transition research: how to account empirically for regime heterogeneity when evaluating the relationship between economic activity, emissions, and policy outcomes. By introducing a regime-diagnostic framework that reconstructs latent transition states prior to empirical modeling, the paper moves beyond pooled analyses that implicitly assume stable and uniform dynamics across countries and time. In doing so, it responds directly to long standing concerns in the Research Policy literature regarding policy learning, comparability, and the reliability of empirical evidence used to inform innovation and transition governance.

Applied to the global $CO_2$-GDP relationship, the analysis shows that climate-economy interactions evolve through distinct transition regimes characterized by different levels of stability, inertia, and predictability. Crucially, these regimes are not merely descriptive labels. They correspond to systematic differences in forecasting accuracy, causal inference stability, and the quality of policy relevant signals extracted from empirical models. As a consequence, identical policy instruments—such as carbon pricing schemes, innovation subsidies, or green industrial strategies—may generate divergent observed outcomes not primarily because of design failures, but because they are deployed under structurally different transition conditions. This finding helps reconcile conflicting results in the empirical literature and reinforces transition theory arguments that policy effectiveness is inherently regime dependent. The main contribution of this paper lies less in proposing new policy instruments than in offering a policy diagnostic perspective that conditions how existing instruments are evaluated and interpreted. By positioning regime identification as a pre-modeling step, the NEED framework complements established econometric and machine learning approaches rather than compet-ing with them. In this sense, NEED functions as a policy evaluation reliability tool: it helps identify when empirical estimates and forecasts are structurally informative, and when they are likely to be distorted by instability and regime shifts. This contribution speaks directly to debates on evidence-based policymaking, adaptive governance, and the limits of average effect analysis in complex socio-technical transitions.

From a broader innovation policy perspective, the findings have implications for policy paradigms and mission-oriented strategies. If policy outcomes depend systematically on transition regimes, then policy design alone is insufficient to ensure effectiveness. Instead, policy learning requires diagnostic capabilities that inform when and under which structural conditions specific instruments are likely to perform as intended. This insight aligns with recent work emphasizing directionality, experimentation, and reflexivity in innovation policy, and suggests that regime-aware diagnostics are a missing link between policy ambitions and empirical evaluation. Beyond its immediate empirical application, the study opens several avenues for future research. First, regime-aware diagnostics can be extended to firm-level, sectoral, or technology-specific data to better connect macro transition dynamics with micro-level innovation processes. Second, integrating regime identification with explicit policy interventions, such as carbon pricing trajectories, mission-oriented programs, or industrial policy packages, would enable more precise analysis of policy timing, sequencing, and interaction effects. Third, comparative applications across domains, including energy, labor, and financial systems, could shed light on how regime dynamics propagate across interconnected socio-technical systems and shape systemic transition risks.

In conclusion, effective innovation and transition policy requires not only appropriate instruments but also a clear understanding of the structural states in which those instruments operate. By making regime heterogeneity empirically visible and analytically operational, this paper contributes to a more context sensitive and diagnostically informed approach to policy evaluation. In doing so, it provides a foundation for future research on adaptive, learning oriented, and regime aware transition governance, an agenda that lies at the core of Research Policy's intellectual mission.

# References

Acemoglu, D., & Restrepo, P. (2018). The race between man and machine: Implications of technology for growth, factor shares, and employment. American Economic Review, 108(6), 1488-1542. https://doi.org/10.1257/aer.20160696

Aghion, P., Dechezleprêtre, A., Hémous, D., Martin, R., & Van Reenen, J. (2016). Carbon taxes, path dependency, and directed technical change: Evidence from the auto industry. Journal of Political Economy, 124(1), 1-51. https://doi.org/10.1086/684581

Angrist, J. D., & Pischke, J. S. (2009). Mostly harmless econometrics: An empiricist's companion. Princeton University Press.

Apergis, N., & Payne, J. E. (2010). Energy consumption and economic growth: Evidence from a panel of OECD countries. Energy Policy, 38(1), 656-660. https://doi.org/10.1016/j.enpol.2009.09.002

Arthur, W. B. (2021). Foundations of complexity economics. Nature Reviews Physics, 3(2), 136-145. https://doi.org/10.1038/s42254-020-00273-3

Arnold, V. I. (1989). Mathematical methods of classical mechanics (2nd ed.). Springer.

Bai, J., & Perron, P. (2003). Computation and analysis of multiple structural change models. Journal of Applied Econometrics, 18(1), 1-22. https://doi.org/10.1002/jae.659

Breiman, L. (2001). Random forests. Machine Learning, 45(1), 5-32. https://doi.org/10.1023/A:1010933404324

Burke, M., Hsiang, S. M., & Miguel, E. (2015). Global non-linear effect of temperature on economic production. Nature, 527(7577), 235-239. https://doi.org/10.1038/nature15725

Chen, K., de Schrijver, E., Sivaraj, S., Sera, F., Scovronick, N., Jiang, L., … Vicedo-Cabrera, A. M. (2024). Impact of population aging on future temperature-related mortality at different global warming levels. Nature Communications, 15(1), 1796. https://doi.org/10.1038/s41467-024-45945-4

Chen, P. (2005). Elasticity and the stability of economic systems. Journal of Economic Behavior & Organization, 56(3), 379-404. https://doi.org/10.1016/j.jebo.2003.09.009

Chen, T., & Guestrin, C. (2016). XGBoost: A scalable tree boosting system. Proceedings of the 22nd ACM SIGKDD Conference, 785-794. https://doi.org/10.1145/2939672.2939785

Dakos, V., Scheffer, M., van Nes, E. H., Brovkin, V., Petoukhov, V., & Held, H. (2008). Slowing down as an early warning signal for abrupt climate change. Proceedings of the National Academy of Sciences, 105(38), 14308-14312. https://doi.org/10.1073/pnas.0802430105

Dakos, V., Carpenter, S. R., van Nes, E. H., & Scheffer, M. (2012). Methods for detecting early warnings of critical transitions in time series illustrated using simulated ecological data. PLoS ONE, 7(7), e41010. https://doi.org/10.1371/journal.pone.0041010

Farmer, J. D., Hepburn, C., Mealy, P., & Teytelboym, A. (2019). A third wave in the economics of climate change. Environmental and Resource Economics, 72(2), 329-357. https://doi.org/10.1007/s10640-018-0272-8

Geels, F. W. (2011). The multi-level perspective on sustainability transitions: Responses to seven criticisms. Environmental Innovation and Societal Transitions, 1(1), 24-40. https://doi.org/10.1016/j.eist.2011.02.002

Grossman, G. M., & Krueger, A. B. (1995). Economic growth and the environment. Quarterly Journal of Economics, 110(2), 353-377. https://doi.org/10.2307/2118443

Hamilton, J. D. (1989). A new approach to the economic analysis of nonstationary time series and the business cycle. Econometrica, 57(2), 357-384. https://doi.org/10.2307/1912559

Hansen, B. E. (2000). Sample splitting and threshold estimation. Econometrica, 68(3), 575-603. https://doi.org/10.1111/1468-0262.00124

Kemp, R., Loorbach, D., & Rotmans, J. (2007). Transition management as a model for managing processes of co-evolution towards sustainable development. International Journal of Sustainable Development & World Ecology, 14(1), 78-91. https://doi.org/10.1080/13504500709469709

Köhler, J., Geels, F. W., Kern, F., Markard, J., Onsongo, E., Wieczorek, A., ... Wells, P. (2019). An agenda for sustainability transitions research: State of the art and future directions. Environmental Innovation and Societal Transitions, 31, 1-32. https://doi.org/10.1016/j.eist.2019.01.004

Lenton, T. M., Livina, V. N., Dakos, V., van Nes, E. H., & Scheffer, M. (2012). Early warning of climate tipping points from critical slowing down. Philosophical Transactions of the Royal Society A, 370(1962), 1185-1204. https://doi.org/10.1098/rsta.2011.0304

Markard, J. (2018). The next phase of the energy transition and its implications for research and policy. Nature Energy, 3(8), 628-633. https://doi.org/10.1038/s41560-018-0171-7

Markard, J., Raven, R., & Truffer, B. (2012). Sustainability transitions: An emerging field of research and its prospects. Research Policy, 41(6), 955-967. https://doi.org/10.1016/j.respol.2012.02.013

Mattioli, G., Roberts, C., Steinberger, J. K., & Brown, A. (2020). The political economy of car dependence. Energy Research & Social Science, 66, 101486. https://doi.org/10.1016/j.erss.2020.101486

Mazzucato, M. (2018). Mission-oriented innovation policies: Challenges and opportunities. Industrial and Corporate Change, 27(5), 803-815. https://doi.org/10.1093/icc/dty034

Peters, G. P., Andrew, R. M., Canadell, J. G., Friedlingstein, P., Jackson, R. B., Korsbakken, J. I., Le Quéré, C., & Peregon, A. (2017). Key indicators to track current progress and future ambition of the Paris Agreement. Nature Climate Change, 7(2), 118-122. https://doi.org/10.1038/nclimate3202

Reichstein, M., Camps-Valls, G., Stevens, B., Jung, M., Denzler, J., Carvalhais, N., & Prabhat. (2019). Deep learning and process understanding for data-driven Earth system science. Nature, 566(7743), 195-204. https://doi.org/10.1038/s41586-019-0912-1

Rogelj, J., Geden, O., Cowie, A., & Reisinger, A. (2021). Three ways to improve net-zero emissions targets. Nature, 591(7850), 365-368. https://doi.org/10.1038/d41586-021-00662-3

Scheffer, M., Bascompte, J., Brock, W. A., Brovkin, V., Carpenter, S. R., Dakos, V., Held, H., van Nes, E. H., Rietkerk, M., & Sugihara, G. (2009). Early-warning signals for critical transitions. Nature, 461(7260), 53-59. https://doi.org/10.1038/nature08227

Schot, J., & Steinmueller, W. E. (2018). Three frames for innovation policy. Research Policy, 47(9), 1554-1567. https://doi.org/10.1016/j.respol.2018.08.011

Stern, D. I. (2004). The rise and fall of the environmental Kuznets curve. World Development, 32(8), 1419-1439. https://doi.org/10.1016/j.worlddev.2004.03.004

Stern, D. I. (2017). The environmental Kuznets curve after 25 years. Journal of Bioeconomics, 19(1), 7-28. https://doi.org/10.1007/s10818-017-9243-1

Unruh, G. C. (2000). Understanding carbon lock-in. Energy Policy, 28(12), 817-830. https://doi.org/10.1016/S0301-4215(00)00070-7

Victor, D. G., Geels, F. W., & Sharpe, S. (2019). Accelerating the low carbon transition. Energy Policy, 132, 1-11. https://doi.org/10.1016/j.enpol.2019.05.026

Wiedmann, T. O., Schandl, H., Lenzen, M., Moran, D., Suh, S., West, J., & Kanemoto, K. (2015). The material footprint of nations. Proceedings of the National Academy of Sciences, 112(20), 6271-6276. https://doi.org/10.1073/pnas.1220362110

# Annex 1: Sample of the study

"Sub-Saharan Africa (SSA)" "Angola", "Benin", "Botswana", "Burkina Faso", "Burundi", "Cameroon", "Cape Verde", "Central African Republic", "Chad", "Comoros", "Congo", "Côte d'Ivoire", "Democratic Republic of the Congo", "Djibouti", "Equatorial Guinea", "Eritrea", "Eswatini", "Ethiopia", "Gabon", "Gambia", "Ghana", "Guinea", "Guinea-Bissau", "Kenya", "Lesotho", "Liberia", "Madagascar", "Malawi", "Mali", "Mauritania", "Mauritius", "Mozambique", "Namibia", "Niger", "Nigeria", "Rwanda", "São Tomé and Príncipe", "Senegal", "Seychelles", "Sierra Leone", "Somalia", "South Africa", "South Sudan", "Sudan", "Tanzania", "Togo", "Uganda", "Zambia", "Zimbabwe"

"Middle East and North Africa (MENA)": [ "Algeria", "Bahrain", "Egypt", "Iran", "Iraq", "Israel", "Jordan", "Kuwait", "Lebanon", "Libya", "Morocco", "Oman", "Palestine", "Qatar", "Saudi Arabia", "Syria", "Tunisia", "Turkey", "United Arab Emirates", "Yemen" ], "Europe and Central Asia": [ "Albania", "Armenia", "Austria", "Azerbaijan", "Belarus", "Belgium", "Bosnia and Herzegovina", "Bulgaria", "Croatia", "Cyprus", "Czech Republic", "Denmark", "Estonia", "Finland", "France", "Georgia", "Germany", "Greece", "Hungary", "Iceland", "Ireland", "Italy", "Kazakhstan", "Kosovo", "Kyrgyzstan", "Latvia", "Lithuania", "Luxembourg", "Malta", "Moldova", "Montenegro", "Netherlands", "North Macedonia", "Norway", "Poland", "Portugal", "Romania", "Russia", "Serbia", "Slovakia", "Slovenia", "Spain", "Sweden", "Switzerland", "Tajikistan", "Turkmenistan", "Ukraine", "United Kingdom", "Uzbekistan" ], "South/East Asia and Pacific": [ "Afghanistan", "Australia", "Bangladesh", "Bhutan", "Brunei", "Cambodia", "China", "Fiji", "India", "Indonesia", "Japan", "Laos", "Malaysia", "Maldives", "Mongolia", "Myanmar", "Nepal", "New Zealand", "North Korea", "Pakistan", "Papua New Guinea", "Philippines", "Samoa", "Singapore", "Solomon Islands", "South Korea", "Sri Lanka", "Taiwan", "Thailand", "Timor-Leste", "Tonga", "Vanuatu", "Vietnam" ], "North America": [ "Canada", "United States", "Mexico" ], "Latin America and Caribbean": [ "Antigua and Barbuda", "Argentina", "Bahamas", "Barbados", "Belize", "Bolivia", "Brazil", "Chile", "Colombia", "Costa Rica", "Cuba", "Dominica", "Dominican Republic", "Ecuador", "El Salvador", "Grenada", "Guatemala", "Guyana", "Haiti", "Honduras", "Jamaica", "Nicaragua", "Panama", "Paraguay", "Peru", "Saint Kitts and Nevis", "Saint Lucia", "Saint Vincent and the Grenadines", "Suriname", "Trinidad and Tobago", "Uruguay", "Venezuela" ]

## Annex 2

Table 1: Understanding the variables used for the study,

| Variable | Meaning in NEED | Interpretation |
|---|---|---|
| Elasticity | Responsiveness of productivity to population changes | Core tension metric; measures system stress between demographic pressure and economic output |
| Velocity | Rate of change of elasticity | Momentum of system adjustment; positive = intensifying tension, negative = relaxing tension |
| Acceleration | Rate of change of velocity | System inertia; measures how rapidly momentum is building or dissipating |
| Jerk | Rate of change of acceleration | System stress; captures abruptness of inertial changes, signaling impending regime shifts |
| Potential Energy | Stored tension from deviation from equilibrium | Accumulated structural imbalance; represents latent pressure for system reorganization |
| Kinetic Energy | Energy from system motion | Active transformation capacity; measures current rate of structural change |
| Hamiltonian | sum of kinetic and potential energy | Overall system vitality; constant in stable regimes, changes indicate energy inflows/outflows |
| Lagrangian | Difference between kinetic and potential energy | System efficiency metric; positive = change-dominated, negative = structure-dominated |
| Acceleration Energy | Energy from changing momentum | Cost of directional changes; high values indicate expensive transitions |
| Jerk Energy | Energy from stress and abruptness | Volatility cost; measures economic "friction" from discontinuous changes |
| Total Energy | Total energy of system | Complete energetic state; diagnostic of regime stability and transition readiness |
| System Power | Rate of energy transformation | System throughput capacity; measures how quickly energy states can change |

# Annex 3

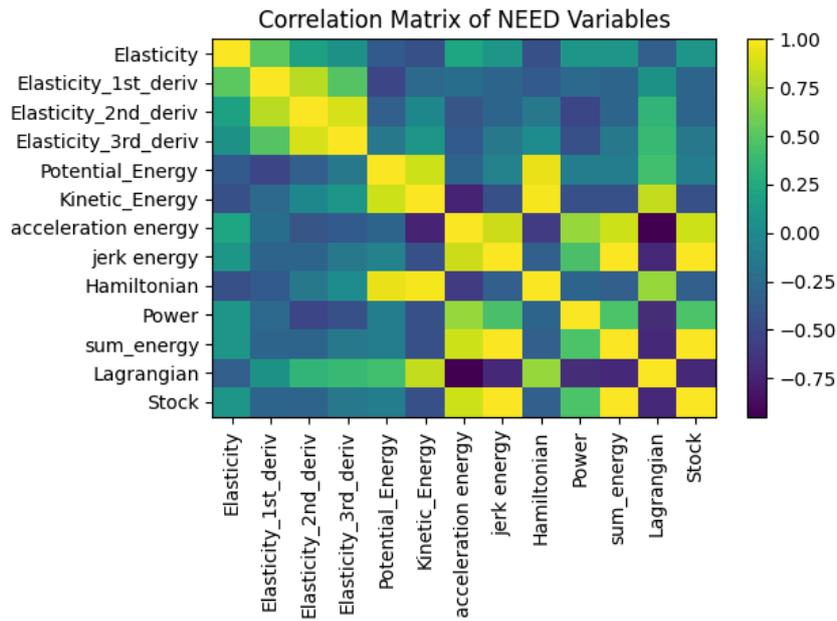

Figure 2: Correlation among variables

# Annex 4

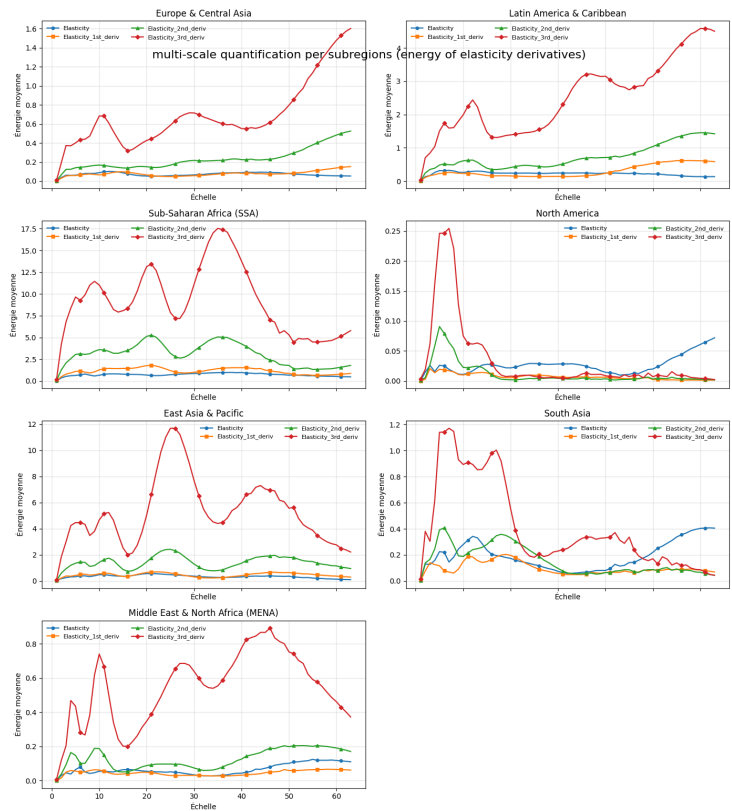

Figure 3: **Store-Power Regimes Visualization**

## Annex 5: Impact of endogenous regimes on prediction of CO2 emission during wealth creation in various subregions

**Table 2: Subregion disparities of NEED Regime Assessment,**

| subregion | Score | Precision | Recall | F1 | AUPRC | Lead time mean | Detect rate | CV Fbeta |
|---|---|---|---|---|---|---|---|---|
| ECA | Kurt5 | 0.149 | 0.929 | 0.257 | 0.149 | 1.013 | 1 | 0.502 |
| MENA | Kurt5 | 0.157 | 0.946 | 0.27 | 0.19 | 1.037 | 0.964 | 0.525 |
| SSA | Kurt5 | 0.149 | 0.988 | 0.258 | 0.168 | 1 | 0.988 | 0.506 |
| LAC | Kurt5 | 0.145 | 0.962 | 0.252 | 0.224 | 1.038 | 1 | 0.502 |
| EAP | Kurt5 | 0.147 | 0.974 | 0.255 | 0.154 | 1 | 1 | 0.507 |
| SA | Kurt5 | 0.148 | 0.857 | 0.253 | 0.316 | 1 | 1 | 0.516 |
| NA | Kurt5 | 0.281 | 0.75 | 0.409 | 0.319 | 1.4 | 0.833 | 0.648 |
| ECA | NN | 0.129 | 0.981 | 0.228 | 0.144 | 1.026 | 1 | 0.469 |
| MENA | NN | 0.136 | 1.0 | 0.239 | 0.227 | 1 | 1 | 0.491 |
| SSA | NN | 0.129 | 0.982 | 0.228 | 0.129 | 1.024 | 1 | 0.471 |
| LAC | NN | 0.125 | 0.923 | 0.221 | 0.187 | 1.154 | 1 | 0.458 |
| EAP | NN | 0.129 | 0.947 | 0.227 | 0.188 | 1.105 | 1 | 0.463 |
| SA | NN | 0.14 | 0.929 | 0.243 | 0.281 | 1.214 | 1 | 0.467 |
| NA | NN | 0.35 | 0.583 | 0.438 | 0.447 | 1 | 0.833 | 0.65 |
| ECA | RF | 0.656 | 0.891 | 0.755 | 0.882 | 1.086 | 0.897 | 0.898 |
| MENA | RF | 0.823 | 0.911 | 0.864 | 0.903 | 1 | 1 | 0.919 |
| SSA | RF | 0.783 | 0.878 | 0.828 | 0.905 | 1.041 | 0.902 | 0.909 |
| LAC | RF | 0.946 | 0.837 | 0.888 | 0.904 | 1 | 0.808 | 0.906 |
| EAP | RF | 0.765 | 0.855 | 0.807 | 0.87 | 1 | 0.842 | 0.903 |
| SA | RF | 0.614 | 0.964 | 0.75 | 0.915 | 1 | 1 | 0.933 |
| NA | RF | 0.706 | 1.0 | 0.828 | 0.975 | 1 | 1 | 1 |
| ECA | Skew5 | 0.179 | 0.891 | 0.299 | 0.231 | 1.068 | 0.949 | 0.555 |
| MENA | Skew5 | 0.157 | 0.946 | 0.27 | 0.269 | 1.036 | 1 | 0.515 |
| SSA | Skew5 | 0.185 | 0.726 | 0.295 | 0.232 | 1.37 | 0.89 | 0.538 |
| LAC | Skew5 | 0.203 | 0.798 | 0.324 | 0.391 | 1.115 | 1 | 0.589 |
| EAP | Skew5 | 0.158 | 0.816 | 0.264 | 0.219 | 1.263 | 1 | 0.505 |
| SA | Skew5 | 0.194 | 0.964 | 0.323 | 0.421 | 1 | 1 | 0.595 |
| NA | Skew5 | 0.175 | 0.583 | 0.269 | 0.423 | 1.667 | 1 | 0.483 |
| ECA | VarCtrl5 | 0.133 | 0.885 | 0.231 | 0.225 | 1.218 | 1 | 0.459 |
| MENA | VarCtrl5 | 0.307 | 0.625 | 0.412 | 0.319 | 1.16 | 0.893 | 0.588 |
| SSA | VarCtrl5 | 0.146 | 0.97 | 0.254 | 0.217 | 1.049 | 1 | 0.494 |
| LAC | VarCtrl5 | 0.234 | 0.673 | 0.347 | 0.248 | 1.745 | 0.904 | 0.501 |
| EAP | VarCtrl5 | 0.188 | 0.737 | 0.299 | 0.22 | 1.441 | 0.895 | 0.507 |
| SA | VarCtrl5 | 0.179 | 0.929 | 0.301 | 0.293 | 1.071 | 1 | 0.536 |
| NA | VarCtrl5 | 0.344 | 0.917 | 0.5 | 0.473 | 1 | 0.833 | 0.613 |
| ECA | XGB | 0.754 | 0.846 | 0.798 | 0.801 | 1.075 | 0.859 | 0.881 |
| MENA | XGB | 0.81 | 0.911 | 0.857 | 0.928 | 1 | 1 | 0.928 |
| SSA | XGB | 0.637 | 0.866 | 0.734 | 0.822 | 1.082 | 0.89 | 0.882 |
| LAC | XGB | 0.778 | 0.875 | 0.824 | 0.872 | 1 | 0.885 | 0.897 |
| EAP | XGB | 0.753 | 0.842 | 0.795 | 0.864 | 1 | 0.842 | 0.876 |
| SA | XGB | 0.833 | 0.893 | 0.862 | 0.881 | 1 | 0.929 | 0.943 |
| NA | XGB | 0.688 | 0.917 | 0.786 | 0.92 | 1 | 0.833 | 0.846 |

**Annex 6:**

Table 4: Prediction performance of NEED within regimes defined exogenously,

| Regime | Model | AUPRC | AUROC | Base rate | PR Lift | Precision | Recall | F1 | Lead time mean | Detect Rate |
| --- | --- | --- | --- | --- | --- | --- | --- | --- | --- | --- |
| Flow | Kurt5 | 0.057 | 0.403 | 0.041 | 1.386 | 0.071 | 0.579 | 0.127 | 1.235 | 0.654 |
| Flow | NN | 0.122 | 0.676 | 0.041 | 2.949 | 0.086 | 0.605 | 0.151 | 1.133 | 0.577 |
| Flow | RF | 0.365 | 0.734 | 0.041 | 8.845 | 0.396 | 0.474 | 0.431 | 1 | 0.423 |
| Flow | Skew5 | 0.06 | 0.438 | 0.041 | 1.452 | 0.074 | 0.566 | 0.131 | 1.353 | 0.654 |
| Flow | VarCtrl5 | 0.147 | 0.672 | 0.041 | 3.562 | 0.074 | 0.645 | 0.132 | 1.176 | 0.654 |
| Flow | XGB | 0.456 | 0.77 | 0.041 | 11.051 | 0.277 | 0.539 | 0.366 | 1.214 | 0.538 |
| Store | Kurt5 | 0.108 | 0.447 | 0.141 | 0.767 | 0.115 | 0.219 | 0.151 | 1.077 | 0.193 |
| Store | NN | 0.191 | 0.573 | 0.141 | 1.353 | 0.15 | 0.977 | 0.26 | 1.225 | 0.889 |
| Store | RF | 0.335 | 0.685 | 0.141 | 2.37 | 0.175 | 0.992 | 0.298 | 1.225 | 0.889 |
| Store | Skew5 | 0.109 | 0.495 | 0.141 | 0.772 | 0.115 | 0.219 | 0.151 | 1.038 | 0.193 |
| Store | VarCtrl5 | 0.19 | 0.51 | 0.141 | 1.342 | 0.111 | 0.212 | 0.146 | 1.037 | 0.2 |
| Store | XGB | 0.309 | 0.688 | 0.141 | 2.188 | 0.174 | 0.985 | 0.296 | 1.242 | 0.889 |
| Transitional | Kurt5 | 0.132 | 0.589 | 0.067 | 1.965 | 0.068 | 0.145 | 0.093 | 1.5 | 0.116 |
| Transitional | NN | 0.071 | 0.467 | 0.067 | 1.046 | 0.066 | 0.927 | 0.123 | 1.811 | 0.768 |
| Transitional | RF | 0.106 | 0.537 | 0.067 | 1.568 | 0.069 | 0.976 | 0.129 | 1.692 | 0.754 |
| Transitional | Skew5 | 0.048 | 0.398 | 0.067 | 0.716 | 0.044 | 0.129 | 0.065 | 2 | 0.116 |